\documentclass[twocolumn,twocolappendix]{aastex631}
\newcommand{\reviseone}{} 
\newcommand{\revisetwo}{} 
\newcommand{\revisethree}{} 
\newcommand{\revisefour}{} 
\newcommand{\revisefive}{} 
\newcommand{\revisesix}{} 

\received{March 13, 2024}
\revised{July 10, 2024}
\accepted{July 30, 2024}

\submitjournal{ApJ}

\shorttitle{Bulk Densities of Small Solar System Bodies} 
\shortauthors{Tatsuuma et al.}

\begin{document}

\title{The Bulk Densities of Small Solar System Bodies as a Probe of Planetesimal Formation}


\correspondingauthor{Misako Tatsuuma}
\email{misako.tatsuuma@gmail.com}

\author[0000-0003-1844-5107]{Misako Tatsuuma}
\affiliation{RIKEN Interdisciplinary Theoretical and Mathematical Science Program (iTHEMS), 2-1 Hirosawa, Wako, Saitama 351-0198, Japan}
\affiliation{Department of Earth and Planetary Sciences, Tokyo Institute of Technology, 2-12-1 Ookayama, Meguro-ku, Tokyo 152-8551, Japan}
\affiliation{Division of Science, National Astronomical Observatory of Japan, 2-21-1 Osawa, Mitaka, Tokyo 181-8588, Japan}

\author[0000-0003-4562-4119]{Akimasa Kataoka}
\affiliation{Division of Science, National Astronomical Observatory of Japan, 2-21-1 Osawa, Mitaka, Tokyo 181-8588, Japan}

\author[0000-0001-9659-658X]{Hidekazu Tanaka}
\affiliation{Astronomical Institute, Graduate School of Science, Tohoku University, 6-3 Aramaki, Aoba-ku, Sendai 980-8578, Japan}

\author[0000-0002-7188-8428]{Tristan Guillot}
\affiliation{Observatoire de la C\^{o}te d’Azur, Boulevard de l’Observatoire CS 34229, F 06304 NICE Cedex 4, France}

\begin{abstract} 

Constraining the formation processes of small solar system bodies is crucial for gaining insights into planetesimal formation.
Their bulk densities, determined by their compressive strengths, offer valuable information about their formation history.
In this paper, we utilize a formulation of the compressive strength of dust aggregates obtained from dust $N$-body simulations to establish the relation between bulk density and diameter.
We find that this relation can be effectively approximated by a polytrope with an index of 0.5, coupled with a formulation of the compressive strength of dust aggregates.
The lowest-density trans-Neptunian objects (TNOs) and main-belt asteroids (MBAs) are well reproduced by dust aggregates composed of 0.1-$\mathrm{\mu}$m-sized grains.
However, most TNOs, MBAs, comets, and near-Earth asteroids (NEAs) exhibit higher densities, suggesting the influence of compaction mechanisms such as collision, dust grain disruption, sintering, or melting, leading to further growth.
We speculate that there are two potential formation paths for small solar system bodies: one involves the direct coagulation of primordial dust grains, resulting in the formation of first-generation planetesimals, including the lowest-density TNOs, MBAs, and parent bodies of comets and NEAs.
In this case, comets and NEAs are fragments or rubble piles of first-generation planetesimals, and objects themselves or rubbles are composed of 0.1-$\mathrm{\mu}$m-sized grains.
The other path involves further potential fragmentation of first-generation planetesimals into compact {\revisesix{dust aggregates observed in protoplanetary disks}}, resulting in the formation of second-generation planetesimals composed of {\revisesix{compact dust aggregates}}, which may contribute to explaining another formation process of comets and NEAs.

\end{abstract}

\keywords{Planet formation (1241), Planetesimals (1259), Small Solar System bodies (1469), Asteroids (72), Comets (280), Near-Earth objects (1092), Trans-Neptunian objects (1705), Main belt asteroids (2036), Comet nuclei (2160), Short period comets (1452), Comet origins (2203), Analytical mathematics (38)}

\section{Introduction} \label{sec:intro}

Constraining the formation \revisetwo{processes} of comets and asteroids in the solar system is crucial for understanding the formation of planetesimals, which are kilometer-sized building blocks of planets.
\revisetwo{The insights into the formation history of comets and asteroids have been provided by some} space explorations, such as Hayabusa \citep[e.g.,][]{Fujiwara2006}, Rosetta \citep[e.g.,][for a review]{Fulle2016,Groussin2019}, Hayabusa2 \citep[e.g.,][]{Sugita2019,Watanabe2019,Kitazato2019}, and \reviseone{Origins, Spectral Interpretation, Resource Identification, Security, Regolith Explorer (OSIRIS-REx)} missions \citep[e.g.,][]{Lauretta2019}.
There are several scenarios for the formation of comets and asteroids: remnants of planetesimals, fragments of planetesimals, or rubble piles, which are bodies made from piles of debris from the collisional destruction of planetesimals \citep[e.g.,][]{Farinella1982,Weissman1986,Weidenschilling1997}.

\reviseone{One of the scenarios for planetesimal formation is the direct coagulation scenario of aggregates of submicrometer-sized dust grains \citep[e.g.,][]{Kataoka2013L}, but it is not consistent with observations of protoplanetary disks \revisetwo{in terms of porosity}.
Some numerical simulations have demonstrated that aggregation of dust grains results in highly porous structures, in other words, the porosity of dust aggregates exceeds $\sim0.99$ \citep[e.g.,][]{Dominik1997,Wada2007,Wada2008,Suyama2008,Suyama2012,Okuzumi2009dustagg,Okuzumi2012,Kataoka2013L}.
\revisetwo{However}, some millimeter polarization observations by the Atacama Large Millimeter/submillimeter Array (ALMA) have indicated that dust aggregates in protoplanetary disks are relatively compact, which means that the porosity is 0.7--0.97 \citep[e.g.,][]{Kirchschlager2019,Tazaki2019,Zhang2023}.}
{\revisesix{Here, we refer to dust aggregates with porosity greater than 0.99 as highly porous dust aggregates, while those with porosity less than 0.99 as compact dust aggregates.}}

Another scenario for planetesimal formation is {\revisefive{the streaming instability \citep[SI, e.g.,][]{Youdin2005} and subsequent gravitational instability \citep[GI, e.g.,][]{Johansen2007}, assuming millimeter--centimeter-sized compact pebbles are the initial condition.
Pebbles are also assumed in}} the pebble accretion scenario \citep[e.g.,][]{Ormel2010,Lambrechts2012}, in which some seed planetesimals capture pebbles to form protoplanets before the dissipation of protoplanetary disk gas.
The {\revisefive{SI/GI formation}} scenario is consistent with some exploration results.
For example, the low tensile strength of comet 67P/Churyumov--Gerasimenko \citep[hereinafter 67P, e.g.,][]{Groussin2015,Basilevsky2016,Hirabayashi2016} is consistent with loosely packed pebble aggregates \citep[e.g.,][]{Skorov2012,Blum2014,Blum2017,Blum2022}.
However, the formation process of compact pebbles from submicrometer-sized dust grains remains unknown.
{\revisesix{Some experiments and numerical simulations have suggested that once dust aggregates become compact with porosity less than $\sim$ 0.99, they bounce and are compressed \citep[e.g.,][]{Weidling2009,Guttler2010,Zsom2010,Krijt2018,Schrapler2022,Arakawa2023}, while highly porous dust aggregates with porosity greater than $\sim$ 0.99 do not bounce \citep[e.g.,][]{Wada2011}.}}

With a scope of constraining the formation process of planetesimals, in this work, we theoretically derive the bulk density of objects made of single-sized grains called monomers and discuss whether it is consistent with the bulk density of small solar system bodies, such as asteroids, comets, and trans-Neptunian objects (TNOs).
We derive a relation between bulk density and diameter of dust aggregates by using polytropes and compressive strength of dust aggregates, which has been investigated numerically \citep[e.g.,][]{Paszun2008,Seizinger2012,Kataoka2013,Tatsuuma2023} and experimentally \citep[e.g.,][]{Blum2004,Guttler2009,Omura2017,Omura2018}.
We use the compressive strength formulated by \citet{Tatsuuma2023}, which can be applied to dust aggregates with volume filling factors both lower and higher than 0.1, because the direct coagulation scenario results in highly porous structures and the subsequent compaction results in less porous structures.
{\revisesix{Moreover, we clarify the dependence of the bulk density on the radius and composition of monomers to trace backward the formation process of planetesimals.
In this work, we assume the direct coagulation scenario resulting in highly porous dust aggregates, which can be considered as a lower limit of the bulk density.}}

This paper is organized as follows.
In Section \ref{sec:method}, we derive \revisefour{a} relation between bulk density and diameter of dust aggregates.
\revisefour{Then}, we \revisefour{compare the relation for dust aggregates with bulk densities and equivalent sphere diameters} of small solar system bodies \revisefour{in Section \ref{sec:results}}.
\revisefour{Based on our results, we discuss the formation processes of small solar system bodies and planetesimals in Sections \ref{subsec:discuss:SSSB} and \ref{subsec:discuss:planetesimal}, respectively.}
We also speculate about the {\revisesix{possible}} origin of {\revisesix{compact dust aggregates observed in protoplanetary disks in terms of porosity}} in Section \ref{subsec:discuss:planetesimal}.
\revisefour{In addition, we note some caveats about the other mechanisms that may affect the bulk density of dust aggregates in Section \ref{subsec:discuss:caveat}.}
Finally, we conclude our work in Section \ref{sec:conclusion}.

\section{\revisefour{Methods}} \label{sec:method}

\revisefour{Here}, we briefly introduce the compressive strength of dust aggregates derived from numerical simulations by \citet{Tatsuuma2023} in Section \ref{subsec:method:comp}.
Then, we \revisefour{derive} the relation between bulk density and diameter in Section \ref{subsec:method:analytic}.
\revisefour{In Section \ref{subsec:method:parameter}, we explain the assumed material parameters of monomers.}

\subsection{Compressive Strength of Dust Aggregates}
\label{subsec:method:comp}

We use a formulation of the compressive strength of dust aggregates \revisefour{with single-sized monomers} by \citet{Tatsuuma2023}, who performed numerical simulations of the compression of dust aggregates.
The compressive strength $P_\mathrm{comp}$ \citep[Equation (11) of][]{Tatsuuma2023} is given \revisetwo{by taking into account the rolling motion of aggregates for low volume filling factors and the closest packing of aggregates for high volume filling factors} as
\begin{eqnarray}
P_\mathrm{comp} &=& \frac{E_\mathrm{roll}}{r_\mathrm{m}^3}\left(\frac{1}{\phi}-\frac{1}{\phi_\mathrm{max}}\right)^{-3}\nonumber\\
&\simeq& 4.7\times10^5\mathrm{\ Pa}\left(\frac{\gamma}{100\mathrm{\ mJ\ m^{-2}}}\right)\left(\frac{r_\mathrm{m}}{0.1\mathrm{\ \mu m}}\right)^{-2}\nonumber\\
&&\times\left(\frac{\xi_\mathrm{crit}}{8\textrm{\ \AA}}\right)\left(\frac{1}{\phi}-\frac{1}{\phi_\mathrm{max}}\right)^{-3},
\label{eq:comp}
\end{eqnarray}
 where $E_\mathrm{roll}=6\pi^2\gamma r_\mathrm{m}\xi_\mathrm{crit}$ is the energy needed for a monomer to roll a distance of $(\pi/2)r_\mathrm{m}$, $r_\mathrm{m}$ is the monomer radius, $\gamma$ is the surface energy of monomers, $\xi_\mathrm{crit}$ is the critical rolling displacement of monomers, $\phi$ is the volume filling factor of dust aggregates, and $\phi_\mathrm{max}=\sqrt{2}\pi/6=0.74$ is the volume filling factor of the closest packing.

Equation (\ref{eq:comp}) shows that dust aggregates are polytropes with a polytropic index {\revisefive{$n$ defined as $P_\mathrm{comp}\propto\rho^{(n+1)/n}$, where $\rho=\rho_\mathrm{m}\phi$ is the density inside dust aggregates and $\rho_\mathrm{m}$ is the material density.
Although the actual index is not constant, $n=0.5$, i.e., $P_\mathrm{comp}\propto\rho^3\propto\phi^3$, if the density is small enough, while $n=0$ if the density is large enough.}}
Polytropes \revisefour{with a} polytropic index \revisefour{of} 0.5 correspond to small dust aggregates, which have small central density and pressure.
On the other hand, polytropes \revisefour{with a} polytropic index \revisefour{of} 0 correspond to large dust aggregates, which have a uniform density and a large pressure.

\subsection{\revisefour{Relation between Bulk Density and Diameter}}
\label{subsec:method:analytic}

We derive a relation between bulk density $\rho_\mathrm{bulk}$ and diameter \reviseone{$D$ based on polytropes \revisefour{with a} polytropic index \revisefour{of} 0.5 \revisefour{and the compressive strength of dust aggregates}.
The diameter $D$ of polytropes with a polytropic index of 0.5 is given by {\revisefive{solving the Lane-Emden equation}} as
\begin{equation}
D = 2\left[\frac{(0.5+1)P_\mathrm{c}}{4\pi G\rho_\mathrm{c}^2}\right]^{1/2}\xi_1
= 2\left(\frac{3P_\mathrm{c}}{8\pi G\rho_\mathrm{c}^2}\right)^{1/2}\xi_1,
\label{eq:polytrope}
\end{equation}
where $P_\mathrm{c}$ is the central pressure\revisethree{, $\rho_\mathrm{c}$ is the central density, $G$ is the gravitational constant,} and \revisetwo{$\xi_1=2.75$} {\revisefive{\citep{Chandrasekhar1939}}}.
In this case of a polytropic index of 0.5, $P_\mathrm{c}/\rho_\mathrm{c}^3 = P_\mathrm{bulk}/\rho_\mathrm{bulk}^3$, where $P_\mathrm{bulk}$ is the bulk pressure.
Therefore,}
\begin{equation}
\frac{P_\mathrm{c}}{\rho_\mathrm{c}^2} = \frac{P_\mathrm{bulk}}{\rho_\mathrm{bulk}^2}\frac{\rho_\mathrm{c}}{\rho_\mathrm{bulk}}
= 1.84\frac{P_\mathrm{bulk}}{\rho_\mathrm{bulk}^2}
\end{equation}
because $\rho_\mathrm{c}/\rho_\mathrm{bulk}=1.84$.
Finally, we obtain
\begin{eqnarray}
D &=& 2\left(\frac{3}{8\pi G}\frac{P_\mathrm{bulk}}{\rho_\mathrm{bulk}^2}\cdot1.84\right)^{1/2}\xi_1 \nonumber\\
&=& 2\left[\frac{3}{8\pi G}\frac{E_\mathrm{roll}}{r_\mathrm{m}^3}\left(\frac{\rho_\mathrm{m}}{\rho_\mathrm{bulk}}-\frac{1}{\phi_\mathrm{max}}\right)^{-3}\frac{1}{\rho_\mathrm{bulk}^2}\cdot1.84\right]^{1/2}\xi_1 \nonumber\\
&\simeq& 5.28\left(\frac{E_\mathrm{roll}}{m_\mathrm{m}G\rho_\mathrm{m}}\right)^{1/2}\left(\frac{\rho_\mathrm{m}}{\rho_\mathrm{bulk}}-\frac{1}{\phi_\mathrm{max}}\right)^{-3/2}\frac{\rho_\mathrm{m}}{\rho_\mathrm{bulk}},
\label{eq:D-rho_bulk}
\end{eqnarray}
where \revisefour{$m_\mathrm{m}=(4/3)\pi r_\mathrm{m}^3\rho_\mathrm{m}$} is the monomer mass, by assuming $P_\mathrm{bulk}=P_\mathrm{comp}$ and using Equation (\ref{eq:comp}).

\revisefour{Equation (\ref{eq:D-rho_bulk}) is the combination of polytropes and dust aggregates because we use Equation (\ref{eq:comp}) instead of a polytropic equation of state.
For the bulk densities comparable to $\rho_\mathrm{m}\phi_\mathrm{max}$, the assumption of the polytropes with an index of 0.5 is not valid.
Therefore, we numerically calculate the internal density and pressure profiles of dust aggregate by solving the hydrostatic equilibrium and compare them with those of polytropes with a constant polytropic index of 0.5 in Appendix \ref{apsec:numerical}.
We confirm that the assumption of a constant index of 0.5 is not valid for high density.
We also confirm that the relation of Equation (\ref{eq:D-rho_bulk}) is a good approximation to that of the exact numerical solution.}

\subsection{\revisefour{Material Parameters}}
\label{subsec:method:parameter}

\revisefour{We assume monomers have material parameters listed in Table \ref{tab:parameters}, and additionally, we assume that each ice monomer has a silicate core whose mass equals the mass of an ice mantle.
The average material density of ice monomers $\rho_\mathrm{m,ice+sil}$ is given as
\begin{equation}
\rho_\mathrm{m,ice+sil}
= \frac{2\rho_\mathrm{m,ice}\rho_\mathrm{m,sil}}{\rho_\mathrm{m,ice}+\rho_\mathrm{m,sil}}
= 1.45\mathrm{\ g\ cm^{-3}},
\end{equation}
because the specific volume of the monomer is $1/\rho_\mathrm{m,ice+sil}=0.5/\rho_\mathrm{m,ice}+0.5/\rho_\mathrm{m,sil}$, where $\rho_\mathrm{m,ice}=1.0\mathrm{\ g\ cm^{-3}}$ and $\rho_\mathrm{m,sil}=2.65\mathrm{\ g\ cm^{-3}}$ are the material densities of ice and silicate.
Hereinafter, we use the material densities of 1.45 and 2.65 $\mathrm{g\ cm^{-3}}$ for ice and silicate monomers, respectively.}

\revisefour{}

\begin{deluxetable}{lcc}
\tablecaption{Material Parameters of Ice and Silicate \label{tab:parameters}}
\tablehead{
\colhead{Parameter} & \colhead{Ice} & \colhead{Silicate}
}
\startdata 
Material density $\rho_\mathrm{m}$ (g cm$^{-3}$) & 1.0 & 2.65 \\
Surface energy $\gamma$ (mJ m$^{-2}$) & 100 & 20 \\
Poisson's ratio $\nu$ & 0.25 & 0.17 \\
Young's modulus $E$ (GPa) & 7 & 54 \\
Critical rolling displacement $\xi_\mathrm{crit}$ (\AA) & 8 & 20 \\
\enddata
\tablecomments{We use $\rho_\mathrm{m}$, $\gamma$, $\nu$, $E$, and $\xi_\mathrm{crit}$ of ice of \citet{Israelachvili1992} and \citet{Dominik1997} and of silicate of \citet{Seizinger2012}.}
\end{deluxetable}

\section{\revisefour{Comparison with Small Solar System Bodies}} \label{sec:results}

In this section, we show our results on the bulk density and diameter of dust aggregates derived by using polytropes \revisefour{with a} polytropic index \revisefour{of} 0.5 \revisefour{and the compressive strength}.
The methods of numerical calculation of solving the hydrostatic equilibrium and the comparison of numerical results with our analytical results are referred to in Appendix \ref{apsec:numerical}.

\revisethree{Figure \ref{fig:size-density} shows the relation between bulk density and diameter of dust aggregates (Equation (\ref{eq:D-rho_bulk})) in four cases: 0.1-$\mathrm{\mu}$m-radius ice, 1.0-$\mathrm{\mu}$m-radius ice, 0.1-$\mathrm{\mu}$m-radius silicate, and 1.0-$\mathrm{\mu}$m-radius silicate monomers.
Here, we refer to primordial dust aggregates as aggregates composed of 0.1--1.0-$\mathrm{\mu m}$-radius monomers.}

We also plot the bulk density and equivalent sphere diameter of small solar system bodies in Figure \ref{fig:size-density}.
The data of TNOs, comets, and asteroids investigated by sample return explorations are listed in Tables \ref{tab:TNOdata}, \ref{tab:cometdata}, and \ref{tab:asteroiddata}, respectively.
\revisethree{We use the data of near-Earth asteroids (NEAs) and main-belt asteroids (MBAs) from Table 1 of \citet{Carry2012} but exclude some data with large errors that exceed the density values for simplicity.}
{\revisefive{Small solar system bodies show trends in their bulk densities and diameters according to their classification.}}
First, TNOs have diameters of 100--1000 km, where bulk densities of dust aggregates converge.
Second, comets have diameters of 1--10 km and bulk densities of 0.1--1 $\mathrm{g\ cm^{-3}}$.
Third, Itokawa, Bennu, Ryugu, and NEAs are smaller and denser than comets, and have diameters of 0.3--3 km and bulk densities of 1--2 $\mathrm{g\ cm^{-3}}$.
Finally, some MBAs are similar to TNOs while MBAs have a large diversity of bulk density.

\revisethree{We also show further growth modes in Figure \ref{fig:size-density}: fragmentation, melting, and monomer disruption.
Fragmentation keeps the bulk density constant but reduces the diameter, while melting and monomer disruption keep the mass constant but reduce the diameter.}
\revisefour{We discuss the possibility of melting and monomer disruption in Section \ref{subsubsec:discuss:caveat:thermal} and Appendix \ref{apsec:monomer}.
Especially, monomer disruption cannot be neglected in the center of dust aggregates due to high pressure, and therefore we derive a relation between central pressure and diameter in Appendix \ref{apsec:central} and the upper limit of diameter below which our results are valid in Appendix \ref{apsec:monomer}.
We find that the upper limit is $\sim 200$ km in the case of 0.1-$\mathrm{\mu m}$-radius monomers, which corresponds to the diameters of TNOs.}

\begin{figure*}[ht!]
\plotone{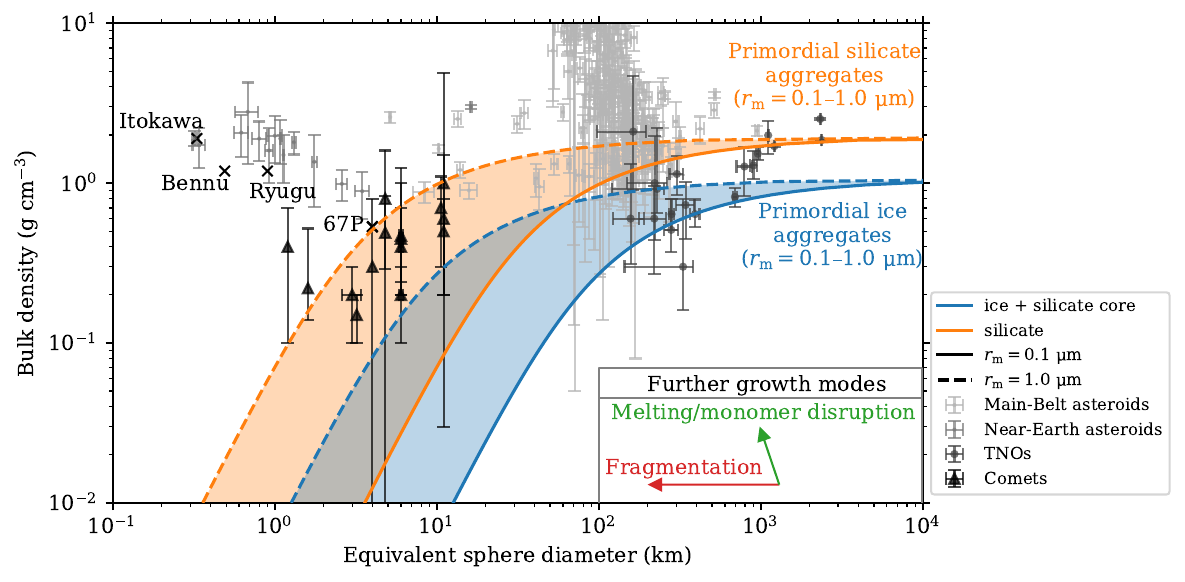} 
\caption{Object's bulk density against equivalent sphere diameter.
\revisethree{All lines show \revisefour{the relation for} dust aggregates (Equation (\ref{eq:D-rho_bulk})).
The blue and orange lines show ice and silicate dust aggregates, whose material densities are 1.45 and 2.65 $\mathrm{g\ cm^{-3}}$, respectively.
The solid and dashed lines show monomer radii of 0.1 and 1.0 $\mathrm{\mu m}$, respectively.
The markers with error bars show MBAs (very light grey dots), NEAs (light grey dots), TNOs (medium grey circles, Table \ref{tab:TNOdata}), and comets (dark grey triangles, Table \ref{tab:cometdata}).}
The black crosses show a comet (67P, Table \ref{tab:cometdata}) and three asteroids (Itokawa, Bennu, and Ryugu, Table \ref{tab:asteroiddata}) which are investigated by in-situ measurement.
\label{fig:size-density}}
\end{figure*}

\begin{deluxetable*}{lccccccl}
\tablecaption{Bulk Density and Diameter of TNOs
\label{tab:TNOdata}}
\tablehead{
\colhead{Name} & \multicolumn{3}{c}{Density ($\mathrm{g\ cm^{-3}}$)} & \multicolumn{3}{c}{Diameter (km)} & \colhead{Reference} \\
\cline{2-4} \cline{5-7}
 &  & \colhead{error} & \colhead{error} &  & \colhead{error} & \colhead{error} & 
}
\startdata 
(134340) Pluto & 1.860 & 0.013 & $-0.013$ & 2374 & 8 & $-8$ & \citet{Stern2015} \\
(136199) Eris & 2.52 & 0.05 & $-0.05$ & 2326 & 12 & $-12$ & \citet{Sicardy2011} \\
(134340) I Charon & 1.702 & 0.021 & $-0.021$ & 1212 & 6 & $-6$ & \citet{Stern2015} \\
(50000) Quaoar & 1.99 & 0.46 & $-0.46$ & 1110 & 5 & $-5$ & \citet{Braga-Ribas2013} \\
(90482) Orcus & 1.53 & 0.15 & $-0.13$ & 958.4 & 22.9 & $-22.9$ & \citet{Fornasier2013} \\
(120347) Salacia & 1.29 & 0.29 & $-0.23$ & 901 & 45 & $-45$ & \citet{Fornasier2013} \\
(174567) Varda & 1.27 & 0.41 & $-0.44$ & 792 & 91 & $-84$ & \citet{Vilenius2014} \\
(55637) 2002 UX$_{25}$ & 0.82 & 0.11 & $-0.11$ & 692 & 23 & $-23$ & \citet{Brown2013} \\
(47171) 1999 TC$_{36}$ & 0.64 & 0.15 & $-0.11$ & 393.1 & 25.2 & $-26.8$ & \citet{Mommert2012} \\
(79360) Sila-Nunam & 0.73 & 0.28 & $-0.28$ & 343 & 42 & $-42$ & \citet{Vilenius2012} \\
(148780) Altjira & 0.30 & 0.50 & $-0.14$ & 331 & 51 & $-187$ & \citet{Vilenius2014} \\
2001 QC$_{298}$ & 1.14 & 0.34 & $-0.30$ & 303 & 27 & $-30$ & \citet{Vilenius2014} \\
(26308) 1998 SM$_{165}$ & 0.51 & 0.29 & $-0.14$ & 279.8 & 29.7 & $-28.6$ & \citet{Stansberry2008,Spencer2006} 
\\
(65489) Ceto & 0.64 & 0.16 & $-0.13$ & 281 & 11 & $-11$ & \citet{Santos-Sanz2012} \\
(275809) 2001 QY$_{297}$ & 0.92 & 1.30 & $-0.27$ & 229 & 22 & $-108$ & \citet{Vilenius2014} \\
2001 XR$_{254}$ & 1.00 & 0.96 & $-0.56$ & 221 & 41 & $-71$ & \citet{Vilenius2014} \\
(88611) Teharonhiawako & 0.60 & 0.36 & $-0.33$ & 220 & 41 & $-44$ & \citet{Vilenius2014} \\
(66652) Borasisi & 2.1 & 2.60 & $-1.2$ & 163 & 32 & $-66$ & \citet{Vilenius2014} \\
(42355) Typhon & 0.60 & 0.72 & $-0.29$ & 157 & 34 & $-34$ & \citet{Stansberry2012} \\
\enddata
\tablecomments{This table is cited from Table A.1. of \citet{Schindler2017}.}
\end{deluxetable*}

\begin{deluxetable*}{lcccllcc}
\tablecaption{
Bulk Density and Equivalent Sphere Diameter of Comets
\label{tab:cometdata}}
\tablehead{
\colhead{Name} & \multicolumn{3}{c}{Density ($\mathrm{g\ cm^{-3}}$)} & \colhead{Method} & \colhead{Reference} & \multicolumn{2}{c}{Diameter (km)} \\
\cline{2-4} \cline{7-8}
 &  & \colhead{lower limit} & \colhead{upper limit} &  &  &  & \colhead{error}
}
\startdata 
1P 
& 0.6 & 0.2 & 1.5 & NGF & \citet{Sagdeev1988} & 11.0 & \nodata \\
& 1 & 0.03 & 4.9 & NGF & \citet{Peale1989} & 11.0 & \nodata \\
& 0.5 & 0.2 & 0.8 & NGF & \citet{Sosa2009} & 11.0 & \nodata \\
2P 
& 0.8 & \nodata & 1.6 & NGF & \citet{Sosa2009} & 4.8 & \nodata \\
6P 
& 0.15 & 0.1 & 0.2 & NGF & \citet{Sosa2009} & 3.2 & \nodata \\
9P 
& 0.4 & 0.2 & 1 & Deep impact ejecta & \citet{Richardson2007} & 6.0 & 0.2 \\
& 0.45 & 0.2 & 0.7 & NGF & \citet{Davidsson2007} & 6.0 & 0.2 \\
& 0.2 & 0.1 & 0.3 & NGF & \citet{Sosa2009} & 6.0 & 0.2 \\
& 0.47 & 0.24 & 1.25 & Deep impact ejecta & \citet{Thomas2013} & 6.0 & 0.2 \\
10P 
& 0.7 & 0.3 & 1.1 & NGF & \citet{Sosa2009} & 10.6 & \nodata \\
19P 
& 0.49 & 0.29 & 0.83 & NGF & \citet{Farnham2002} & 4.8 & \nodata \\
22P 
& 0.2 & 0.1 & 0.3 & NGF & \citet{Sosa2009} & 3.0 & 0.4 \\
46P 
& 0.4 & 0.1 & 0.7 & NGF & \citet{Sosa2009} & 1.2 & \nodata \\
67P 
& 0.535 & 0.5 & 0.57 & In-situ mass and volume & \citet{Preusker2015} & 4.0 & \nodata \\
& 0.532 & 0.525 & 0.539 & In-situ mass and volume & \citet{Jorda2016} & 4.0 & \nodata \\
81P 
& 0.3 & \nodata & 0.8 & NGF & \citet{Sosa2009} & 4.0 & \nodata \\
103P 
& 0.22 & 0.14 & 0.52 & Erosion rate & \citet{Richardson2014} & 1.6 & \nodata \\
\enddata
\tablecomments{We exclude the data without the preferred value of density in \citet{Groussin2019}.
The reference of diameters is \url{https://ssd.jpl.nasa.gov/sbdb_query.cgi}.
NGF denotes non-gravitational forces, which cause an acceleration of a comet's nucleus and change its orbital parameters, so that its mass can be estimated.}
\end{deluxetable*}

\begin{deluxetable*}{lcccccl}
\tablecaption{Bulk Density and Equivalent Sphere Diameter of Asteroids Investigated by Sample Return Explorations
\label{tab:asteroiddata}}
\tablehead{
\colhead{Name} & \multicolumn{2}{c}{Density ($\mathrm{g\ cm^{-3}}$)} & \multicolumn{2}{c}{Volume (km$^3$)} & \colhead{Diameter (km)} & \colhead{Reference} \\
\cline{2-3} \cline{4-5}
 &  & \colhead{error} &  & \colhead{error} &  & 
}
\startdata 
162173 Ryugu & 1.19 & 0.02 & 0.377 & 0.0049 & 0.90 & \citet{Watanabe2019} \\
101955 Bennu & 1.19 & 0.013 & 0.0615 & 0.0001 & 0.49 & \citet{Lauretta2019} \\
25143 Itokawa & 1.9 & 0.13 & 0.0184 & 0.00092 & 0.33 & \citet{Fujiwara2006} \\
\enddata
\tablecomments{We derive their diameters from their volumes by assuming that they are spheres.}
\end{deluxetable*}


\section{\revisefour{Formation of Small Solar System Bodies}}
\label{subsec:discuss:SSSB}

In this section, we compare our results with those of \revisethree{TNOs, comets, and asteroids} plotted in Figure \ref{fig:size-density}, and discuss their formation processes in Sections \revisethree{\ref{subsubsec:discuss:SSSB:TNO}, \ref{subsubsec:discuss:SSSB:comet}, and \ref{subsubsec:discuss:SSSB:asteroid}, respectively.
We assume some formation scenarios of small solar system bodies: primordial dust aggregates or further growth modes.
The further growth modes contain fragmentation and compaction; fragmentation means fragments or rubble piles of primordial dust aggregates and compaction means additionally compacted dust aggregates due to melting or monomer disruption.}

\subsection{Formation Process of TNOs}
\label{subsubsec:discuss:SSSB:TNO}

\revisethree{We can explain TNOs, which have diameters of 100--1000 km and bulk densities of 0.3--2 $\mathrm{g\ cm^{-3}}$, in terms of primordial dust aggregates and additionally compacted dust aggregates.}

\revisethree{Figure \ref{fig:size-density} shows that the lowest density TNO can be explained by primordial ice aggregates composed of 0.1-$\mathrm{\mu m}$-sized monomers.
The diversity of TNOs' bulk densities can be explained by the diversity of monomer radius, although its upper limit cannot be constrained because bulk densities of dust aggregates converge in the diameter range of TNOs.
The diversity of composition can also explain the diversity of TNOs' bulk densities.
For further discussion about the composition of TNOs, see other works, for example, \citet{Bierson2019}.}

\revisethree{The additionally compacted dust aggregates due to melting or monomer disruption can explain the diversity of TNOs' bulk densities.
TNO-sized objects experience thermal evolution due to the decay energy of radionuclides (Section \ref{subsubsec:discuss:caveat:thermal}).
In addition, monomers in the center of TNO-sized objects can be broken due to high central pressure \revisefour{(Appendices \ref{apsec:central} and \ref{apsec:monomer})}.}

\subsection{Formation Process of Comets}
\label{subsubsec:discuss:SSSB:comet}

\revisethree{We can explain comets, which have diameters of 1--10 km and bulk densities of 0.1--1 $\mathrm{g\ cm^{-3}}$, in terms of fragments or rubble piles of primordial dust aggregates and additionally compacted dust aggregates.}

\revisethree{Figure \ref{fig:size-density} shows that even the lowest-density comet cannot be explained by primordial ice aggregates composed of 0.1--1.0 $\mathrm{\mu m}$-sized monomers.
If comets are primordial dust aggregates, the monomer radius should be} $1.0\mathrm{\ \mu m}<r_\mathrm{m}\lesssim10.0\mathrm{\ \mu m}$ for ice and $0.3\mathrm{\ \mu m}<r_\mathrm{m}<3.0\mathrm{\ \mu m}$ for silicate.
Especially, we estimate the monomer radius of comet 67P, whose bulk density is precisely estimated, by using
\begin{equation}
r_\mathrm{m} \simeq 19.8\left(\frac{\gamma\xi_\mathrm{crit}}{\rho_\mathrm{m}^2 D^2 G}\right)^{1/2}\left(\frac{\rho_\mathrm{m}}{\rho_\mathrm{bulk}}-\frac{1}{\phi_\mathrm{max}}\right)^{-3/2}\frac{\rho_\mathrm{m}}{\rho_\mathrm{bulk}}
\label{eq:r_m-D-rho_bulk}
\end{equation}
from Equation (\ref{eq:D-rho_bulk}).
If we assume that $\rho_\mathrm{bulk}=0.532\mathrm{\ g\ cm^{-3}}$ \citep{Jorda2016} and $D=4$ km, we obtain \reviseone{$r_\mathrm{m}\simeq6.3\mathrm{\ \mu m}$} for ice and $r_\mathrm{m}\simeq1.0\mathrm{\ \mu m}$ for silicate by using Equation (\ref{eq:r_m-D-rho_bulk}).
However, the derived monomer radius is not consistent with the monomer radius that can explain comets' low tensile strength, and therefore we may have to consider other mechanisms that make strength low.
For example, \citet{Tatsuuma2019} have shown that the tensile strength of dust aggregates composed of submicrometer-sized monomers is higher than the estimated tensile strength of comets.
Especially, they have suggested that comet 67P is composed of 3.3--220-$\mathrm{\mu m}$-sized silicate monomers, while our results suggest 1.0-$\mathrm{\mu m}$ silicate monomers.
This inconsistency may be caused by other missing mechanisms to explain the low tensile strength of comets, such as sintering \citep[e.g.,][]{Sirono2017} and the dependence of tensile strength on the volume of dust aggregates \citep{Kimura2020}.
We note that the compressive and tensile strengths of pebble aggregates {\revisefive{formed via SI/GI}} can be different from those of dust aggregates if comets are composed of compact pebbles {\revisefive{\citep[e.g.,][]{Skorov2012}}}, but investigating the strength of pebble aggregates {\revisefive{numerically}} is future work.

\revisethree{Figure \ref{fig:size-density} suggests that the comet formation process needs further growth modes: fragmentation and compaction.
Fragmentation means fragments or rubble piles and needs parent dust aggregates.
If we do not assume compaction, we can estimate that the diameters of parent dust aggregates} are \revisethree{$\sim60$ km or larger} for 0.1-$\mathrm{\mu m}$-sized ice monomers, \revisethree{larger than 10 km} for 0.1-$\mathrm{\mu m}$-sized silicate monomers, and \revisethree{$\sim6$ km or larger} for 1.0-$\mathrm{\mu m}$-sized ice monomers.
Especially, we can estimate the diameter of the parent body of comet 67P.
If we assume that $\rho_\mathrm{bulk}=0.532\mathrm{\ g\ cm^{-3}}$ \citep{Jorda2016}, we obtain $D\simeq2.5\times10^2$ km for 0.1-$\mathrm{\mu m}$-sized ice monomers, $D\simeq42$ km for 0.1-$\mathrm{\mu m}$-sized silicate monomers, and $D\simeq25$ km for 1.0-$\mathrm{\mu m}$-sized ice monomers by using Equation (\ref{eq:D-rho_bulk}).
However, the diameters of parent \revisethree{dust aggregates} are only approximate values, and some mechanisms may increase and decrease the bulk densities of comets.
Their bulk densities become lower than expected if comets are composed of surface materials of the parent \revisethree{dust aggregates}, while their bulk densities become higher than expected if comets are composed of central materials of the parent \revisethree{dust aggregates}.
Moreover, melting \revisethree{and monomer disruption} inside the parent \revisethree{dust aggregates and the collisional compression} make the bulk densities of comets higher, while voids among the rubble make the bulk densities of comets lower if comets are rubble piles.
{\revisefive{We note that the tensile strength of dust aggregates is higher than that of comets, and therefore we need to consider sintering \citep[e.g.,][]{Sirono2017} and volume dependence \citep{Kimura2020} that lowers the tensile strength of dust aggregates if comets are single fragments, as well as possible pebble aggregates formed via SI/GI \citep[e.g.,][]{Skorov2012}.}}

\subsection{Formation Process of Asteroids}
\label{subsubsec:discuss:SSSB:asteroid}

\revisethree{We can explain the majority of MBAs, which have diameters of $\sim100$ km and bulk densities of $\gtrsim 1\mathrm{\ g\ cm^{-3}}$, in terms of primordial dust aggregates and additionally compacted dust aggregates, while we can explain Itokawa, Bennu, Ryugu, and NEAs, which have diameters of 0.3--3 km and bulk densities of 1--2 $\mathrm{\ g\ cm^{-3}}$, in terms of fragments or rubble piles of primordial dust aggregates and additionally compacted dust aggregates.}

\revisethree{Figure \ref{fig:size-density} shows that some MBAs can be explained by primordial silicate aggregates composed of 0.1--1.0 $\mathrm{\mu m}$-sized monomers.
The dispersion of MBAs' bulk densities is much larger than that due to different monomer radii, and therefore we need further growth modes.}

\revisethree{The additionally compacted dust aggregates due to melting or monomer disruption can explain the dispersion of MBAs' bulk densities.
100-km-sized objects experience thermal evolution due to the decay energy of radionuclides, such as melting and differentiation (Section \ref{subsubsec:discuss:caveat:thermal}).
In addition, monomers in the center of objects whose diameter exceeds $\simeq 200$ km can be broken due to high central pressure \revisefour{(Appendices \ref{apsec:central} and \ref{apsec:monomer})}.}

\revisethree{In the case of Itokawa, Bennu, Ryugu, and NEAs, Figure \ref{fig:size-density} shows that they cannot be explained by primordial silicate aggregates composed of 0.1--1.0 $\mathrm{\mu m}$-sized monomers.
If these asteroids are primordial silicate aggregates, the monomer radius should be $r_\mathrm{m}\gtrsim10.0\mathrm{\ \mu m}$}, while the bulk density of Itokawa is too high to be explained by \revisethree{primordial silicate} aggregates, which suggests its parent body has experienced melting to make it denser.
\revisethree{Especially,} we estimate the monomer radius of Ryugu and Bennu by using their bulk densities and diameters in Table \ref{tab:asteroiddata}\revisethree{, and we} obtain $r_\mathrm{m}\simeq18\mathrm{\ \mu m}$ for Ryugu and \reviseone{$r_\mathrm{m}\simeq32\mathrm{\ \mu m}$} for Bennu by using Equation \reviseone{(\ref{eq:r_m-D-rho_bulk})}.

\revisethree{Figure \ref{fig:size-density} suggests that the formation processes of Itokawa, Bennu, Ryugu, and NEAs need further growth modes: fragmentation and compaction.
If we do not assume compaction, we can estimate that the diameters of parent dust aggregates are $\sim1.0\times10^2$ km or larger for 0.1-$\mathrm{\mu m}$-sized silicate monomers and $\sim10$ km or larger for 1.0-$\mathrm{\mu m}$-sized silicate monomers.
Especially, we can estimate the diameter of the parent bodies of Ryugu and Bennu, while we cannot estimate that of Itokawa because of possible melting.}
If we assume that $\rho_\mathrm{bulk}=1.19\mathrm{\ g\ cm^{-3}}$ for Ryugu and Bennu \citep{Watanabe2019,Lauretta2019}, we obtain $D\simeq1.6\times10^2$ km for 0.1-$\mathrm{\mu m}$-sized silicate monomers and $D\simeq16$ km for 1.0-$\mathrm{\mu m}$-sized silicate monomers by using Equation (\ref{eq:D-rho_bulk}).
The \revisethree{previously} suggested size of the parent body of Ryugu, which is 100--160 km \citep{Walsh2013}, is consistent with our results of 0.1-$\mathrm{\mu m}$-sized silicate monomers.
\revisethree{The} parent bodies of Ryugu are believed to be Eulalia and Polana from the results of the observed visible spectral type and orbital dynamics \citep[e.g.,][]{Sugita2019}.
Eulalia is a member of the Eulalia family, whose member asteroids are dynamically linked.
The Eulalia family is thought to be formed as a result of the breakup of a 100--160 km parent body \citep{Walsh2013}, which is consistent with our results that $D\gtrsim1.6\times10^2$ km for 0.1-$\mathrm{\mu m}$-sized silicate monomers.
However, our results of the parent bodies' diameters have some uncertainty \revisethree{because of some mechanisms that increase and decrease the bulk densities of fragments (see Section \ref{subsubsec:discuss:SSSB:comet})}, and therefore we cannot exclude the possibility of 1.0-$\mathrm{\mu m}$-sized silicate monomers.

\section{Planetesimal Formation and Origin of {\revisesix{Observed Compact Dust Aggregates}}}
\label{subsec:discuss:planetesimal}

In this section, we discuss the formation process of planetesimals and the origin of compact {\revisesix{dust aggregates}} observed in protoplanetary disks.
First, we speculate the formation process of {\revisesix{observed compact dust aggregates}} in terms of porosity in Section \ref{subsubsec:discuss:planetesimal:pebble}.
Then, we propose a unified formation scenario of planetesimals including \revisethree{TNOs,} comets, asteroids, and {\revisesix{observed compact dust aggregates}} in terms of direct coagulation in Section \ref{subsubsec:discuss:planetesimal:planetesimal}.

\subsection{Formation Process of {\revisesix{Observed Compact Dust Aggregates}}}
\label{subsubsec:discuss:planetesimal:pebble}

First, we focus on millimeter-sized {\revisesix{compact dust aggregates observed in protoplanetary disks}}, whose porosity is 0.7--0.97 \citep[e.g.,][]{Kirchschlager2019,Tazaki2019,Zhang2023}.
The bulk density of compact {\revisesix{dust aggregates}} is 0.044--0.44 {\revisefive{$\mathrm{g\ cm^{-3}}$}} if we assume that {\revisesix{they}} are composed of ice with silicate cores, whose material density is 1.45 $\mathrm{g\ cm^{-3}}$.

Figure \ref{fig:size-density} suggests that compact {\revisesix{dust aggregates}} need their parent {\revisesix{bodies}} to achieve the bulk density of {\revisesix{observed compact dust aggregates}} and {\revisesix{they}} are fragments of parent {\revisesix{bodies}}.
Equation (\ref{eq:D-rho_bulk}) shows that their diameters are 30--180 km or larger for 0.1-$\mathrm{\mu m}$-sized ice monomers and 3--18 km or larger for 1.0-$\mathrm{\mu m}$-sized ice monomers.
{\revisesix{Compression mechanisms other than self-gravitational compression cannot compress highly porous dust aggregates with porosity greater than 0.99, which is expected from direct coagulation,}} as \citet{Kataoka2013L} have shown that the bulk density of dust aggregates compressed by collisions and gas pressure is {\revisefive{from $10^{-5} \mathrm{\ g\ cm^{-3}}$ to $10^{-3} \mathrm{\ g\ cm^{-3}}$, which is}} much smaller than that determined by self-gravitational compression.

\subsection{Speculation about Planetesimal Formation Process}
\label{subsubsec:discuss:planetesimal:planetesimal}

\revisethree{Next}, we discuss and propose a unified formation scenario of planetesimals including \revisethree{TNOs,} comets, asteroids, and {\revisesix{observed compact dust aggregates}} in terms of direct coagulation.
First, we summarize the possible formation process of \revisethree{TNOs,} comets, asteroids, and {\revisesix{observed compact dust aggregates}} in Table \ref{tab:formationprocess} based on our results in Sections \ref{subsubsec:discuss:SSSB:TNO}, \ref{subsubsec:discuss:SSSB:comet}, \ref{subsubsec:discuss:SSSB:asteroid}, and \ref{subsubsec:discuss:planetesimal:pebble}.

We can constrain the monomer radius as an initial condition of planet formation to 0.1 $\mathrm{\mu m}$ \revisethree{for both ice and silicate} by the discussion about \revisethree{the monomer radius of the lowest density TNO and the diameter of} the parent body of Ryugu in \revisethree{Sections \ref{subsubsec:discuss:SSSB:TNO} and \ref{subsubsec:discuss:SSSB:asteroid}, respectively.}
\revisethree{Once we can constrain the monomer radius to 0.1 $\mathrm{\mu m}$ and if we assume the same monomer radius as that of comets, there are two formation \revisefour{scenarios} for NEAs and comets: direct coagulation and fragments of primordial dust aggregates, or aggregation of {\revisesix{compact dust aggregates}}.}

\begin{figure*}[ht!]
\plotone{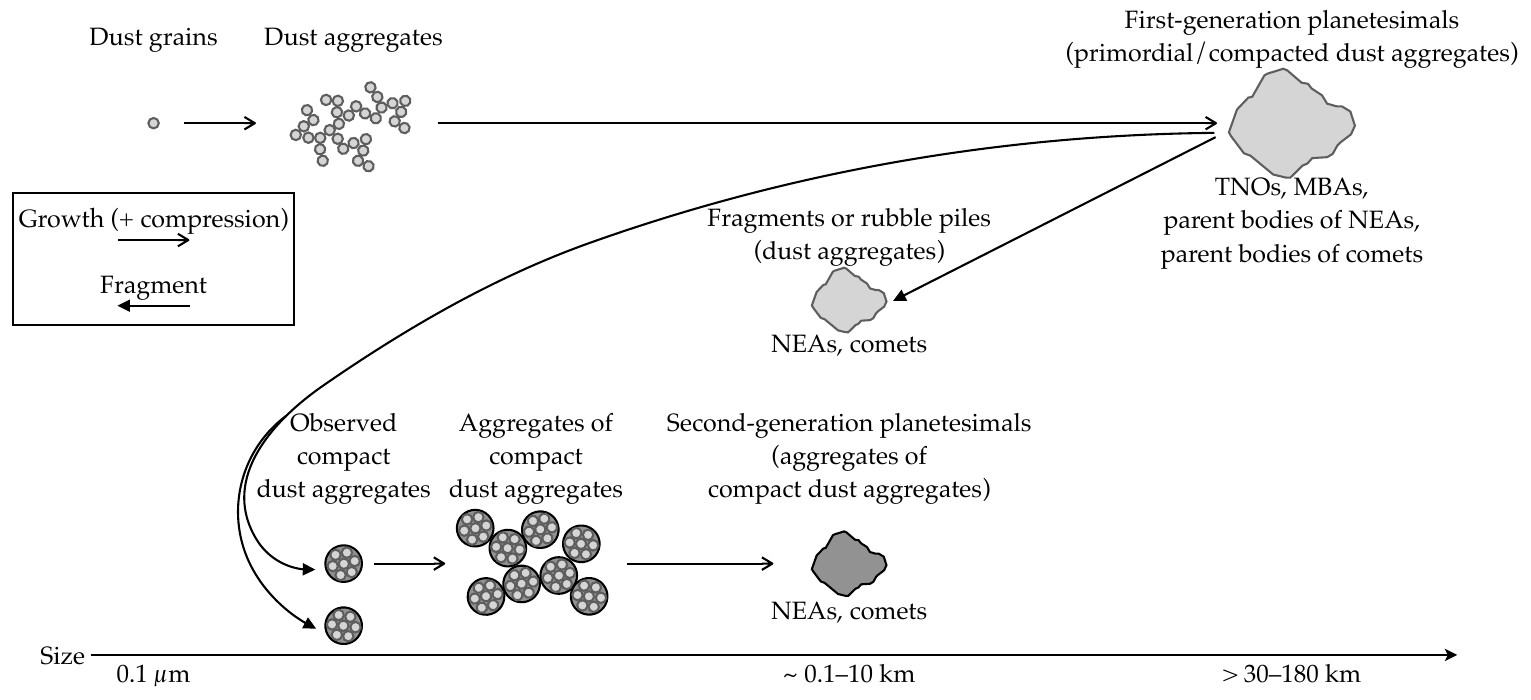} 
\caption{The proposed unified formation scenario of planetesimals including \revisethree{TNOs,} comets, asteroids, and {\revisesix{observed compact dust aggregates}}.
First, \revisethree{0.1-$\mathrm{\mu m}$}-sized dust grains coagulate and grow into dust aggregates, and then the first-generation planetesimals composed of dust aggregates form.
The first-generation planetesimals include \revisethree{TNOs, MBAs, parent bodies of NEAs including Ryugu and Bennu, and parent bodies of comets.}
\revisefour{In this case, NEAs and comets are fragments or rubble piles of the first-generation planetesimals.}
Second, \revisethree{a part of} first-generation planetesimals fragment into {\revisesix{observed compact dust aggregates}}.
Then, {\revisesix{compact dust aggregates}} coagulate and grow into {\revisesix{aggregates of compact dust aggregates}}, and then the second-generation planetesimals composed of {\revisesix{aggregates of compact dust aggregates}} form.
The second-generation planetesimals \revisethree{include NEAs and comets}.
\label{fig:formation_scenario}}
\end{figure*}

\revisethree{The first formation \revisefour{scenario} needs primordial dust aggregates with a diameter of 30--180 km or larger, which we call the first-generation planetesimals, as shown in Figure \ref{fig:formation_scenario}.
The first-generation planetesimals can experience compaction due to melting or monomer disruption because of their diameter.
The diameter of the first-generation planetesimals is consistent with diameters of TNOs, MBAs, parent bodies of NEAs including Ryugu and Bennu, parent bodies of comets, and parent bodies of {\revisesix{observed compact dust aggregates}} as shown in Table \ref{tab:formationprocess}.
In this case, NEAs and comets are fragments or rubble piles, and {\revisesix{observed compact dust aggregates}} are fragments of planetesimals.}
Our results suggest that there are planetesimals in protoplanetary disks where compact {\revisesix{dust aggregates}} are observed.

\revisethree{The second formation \revisefour{scenario} needs planetesimals composed of {\revisesix{aggregates of compact dust aggregates}}, which we call the second-generation planetesimals, because there is a contradiction between the expected tensile strength of dust aggregates composed of 0.1-$\mathrm{\mu m}$-radius monomers and the estimated tensile strength of comets as discussed in Section \ref{subsubsec:discuss:SSSB:comet}.
One of the possible explanations for the low tensile strength of comets is {\revisesix{aggregates of compact dust aggregates}}.}
Our results suggest that there is a possibility that \revisethree{NEAs and} comets are second-generation planetesimals whose diameter is \revisethree{0.1--10} km as shown in Figure \ref{fig:formation_scenario}.


\begin{deluxetable*}{lcccccc}
\tablecaption{Possible Formation Process of \revisethree{TNOs,} Comets, Asteroids, and {\revisesix{Observed compact dust aggregates}}
\label{tab:formationprocess}}
\tablehead{
\colhead{Objects} & \multicolumn{2}{c}{\revisethree{Primordial dust aggregates}} & \multicolumn{4}{c}{Fragments or rubble piles} \\
\cline{2-3} \cline{4-7}
 & \colhead{Ice} & \colhead{Silicate} & \multicolumn{2}{c}{Ice} & \multicolumn{2}{c}{Silicate} \\
\cline{4-5} \cline{6-7}
 & & & \colhead{$r_\mathrm{m}=0.1\mathrm{\ \mu m}$} & \colhead{$r_\mathrm{m}=1.0\mathrm{\ \mu m}$} & \colhead{$r_\mathrm{m}=0.1\mathrm{\ \mu m}$} & \colhead{$r_\mathrm{m}=1.0\mathrm{\ \mu m}$}
}
\startdata 
\revisethree{TNOs} & \revisethree{$r_\mathrm{m}\gtrsim0.1\mathrm{\ \mu m}$} & \nodata & \nodata & \nodata & \nodata & \nodata \\
Comets & $1.0\mathrm{\ \mu m}<r_\mathrm{m}\lesssim10.0\mathrm{\ \mu m}$ & $0.3\mathrm{\ \mu m}<r_\mathrm{m}<3.0\mathrm{\ \mu m}$ & $D\gtrsim60$ km & $D\gtrsim6$ km & $D>10$ km & \nodata \\
\revisethree{67P} & \revisethree{$r_\mathrm{m}\simeq6.3\mathrm{\ \mu m}$} & \revisethree{$r_\mathrm{m}\simeq1.0\mathrm{\ \mu m}$} & \revisethree{$D\simeq2.5\times10^2$ km} & \revisethree{$D\simeq25$ km} & \revisethree{$D\simeq42$ km} & \nodata \\
\revisethree{MBAs} & \nodata & \revisethree{$r_\mathrm{m}\gtrsim0.1\mathrm{\ \mu m}$} & \nodata & \nodata & \nodata & \nodata \\
\revisethree{NEAs} & \nodata & \revisethree{$r_\mathrm{m}\gtrsim10.0\mathrm{\ \mu m}$} & \nodata & \nodata & \revisethree{$D\gtrsim1.0\times10^2$ km} & \revisethree{$D\gtrsim10$ km} \\
\revisethree{Ryugu} & \nodata & \revisethree{$r_\mathrm{m}\simeq18\mathrm{\ \mu m}$} & \nodata & \nodata & \revisethree{$D\simeq1.6\times10^2$ km} & \revisethree{$D\simeq16$ km} \\
\revisethree{Bennu} & \nodata & \revisethree{$r_\mathrm{m}\simeq32\mathrm{\ \mu m}$} & \nodata & \nodata & \revisethree{$D\simeq1.6\times10^2$ km} & \revisethree{$D\simeq16$ km} \\
{\revisesix{Observed}}\\
{\revisesix{compact}} & \nodata & \nodata & \revisethree{$D\gtrsim30$--180} km & \revisethree{$D\gtrsim3$--18} km & \nodata & \nodata \\
{\revisesix{dust aggregates}}
\enddata
\tablecomments{$r_\mathrm{m}$ is the monomer radius and $D$ is the diameter of the parent body.}
\end{deluxetable*}

\section{\revisefour{Caveats}}
\label{subsec:discuss:caveat}

In this section, we discuss some caveats about the other mechanisms that may affect the bulk density of dust aggregates\revisethree{: thermal evolution of km-sized objects, uncertainty of material parameters, and pebble aggregates in Sections \ref{subsubsec:discuss:caveat:thermal}, \ref{subsubsec:discuss:caveat:uncertaintyparam}, and \ref{subsubsec:discuss:caveat:pebbleaggregate}, respectively.}

\subsection{Thermal Evolution of km-sized Objects}
\label{subsubsec:discuss:caveat:thermal}

\revisethree{We discuss the possibility of melting of monomers by the head-producing short-lived radionuclides, such as $^{26}$Al \citep[e.g.,][]{Urey1955}.
If the temperature exceeds $\sim1400$ K, silicate would start melting \citep{Agee1995} and our calculations are no longer valid.
Moreover, if the temperature exceeds $\sim300$ K, ice would start melting although the temperature depends on the pressure.
Here, we roughly discuss at which diameter the maximum temperature may reach these limits and monomers cannot keep their shape.} 

\revisethree{The previous calculations of temperature increase inside planetesimals have normally assumed fixed diameter, porosity, thermal conductivity, and formation timing, which determine the amount of $^{26}$Al included \citep[e.g.,][for a review]{Gail2014}.
For example, \citet{Henke2012} estimated the maximum temperature that can be reached inside planetesimals.
If the porosity is 0 and the formation time is the same as that of calcium-aluminium-rich inclusions (CAIs), which is the most heated case, the planetesimals with a diameter of $\simeq16$ km reach 1500 K, and those with a diameter of $\simeq4$ km reach 400 K while smaller bodies do not reach such temperatures.
If the porosity is 0.5, the planetesimals with a diameter of $\simeq2$ km reach 1500 K, and those with a diameter of sub-km reach 400K.
We note that the small bodies we treat in this paper can experience melting, although the porosity dependence of the maximum temperature is not fully understood.}

\subsection{\revisefour{Uncertainty of Material Parameters}}
\label{subsubsec:discuss:caveat:uncertaintyparam}

Equations (\ref{eq:D-rho_bulk}) and (\ref{eq:r_m-D-rho_bulk}) show the relation among the object's diameter, object's bulk density, and monomer radius, and demonstrate that they depend on some material parameters, which may have uncertainty.
For example, the critical rolling displacement of monomers $\xi_\mathrm{crit}$ in $E_\mathrm{roll}$ has a range from 2 {\AA} of theory \citep{Dominik1997} to 32 {\AA} of experiments \citep{Heim1999}.
The surface energy of monomers $\gamma$ in $E_\mathrm{roll}$ also has uncertainty \citep[e.g.,][]{Yamamoto2014,Kimura2015,Gundlach2015,Gundlach2018} depending on materials and temperatures.
Moreover, the material density has uncertainty depending on the mass ratio between an ice mantle and a silicate core, although we assume 1:1 in this work.
If we assume other materials, such as carbon, the material density, surface energy, and other parameters can be different from the values of ice and silicate.
However, the object's diameter and bulk density of other materials should be between those of ice and silicate if the material parameters have values between those of ice and silicate.

\subsection{Bulk Density of Pebble Aggregates}
\label{subsubsec:discuss:caveat:pebbleaggregate}

Pebble aggregates may have a different relation between their bulk density and diameter, which is beyond the scope of this work.
Their compressive strength, which determines their bulk density, may be different from that of dust aggregates in this work and can be more fragile than them \citep[e.g.,][]{Skorov2012}.
However, the compressive strength of pebble aggregates has not been investigated by numerical simulations.
The bulk density of pebble aggregates can be higher if their compressive strength is lower than that of dust aggregates, while the bulk density of pebble aggregates tends to be lower because they have voids both inside and between pebbles.

\section{Conclusions} \label{sec:conclusion}

We \revisethree{derived the relation between bulk density and diameter of dust aggregates by using polytropes and compressive strength formulated by \citet{Tatsuuma2023}.}
Then, we compared our results with those of small solar system bodies to \revisethree{discuss} their formation process as well as the planetesimal formation process.

Our main findings are as follows.

\begin{enumerate}

\item The relation between bulk density and diameter of dust aggregates is given by Equation (\ref{eq:D-rho_bulk}) by using polytropes \revisefour{with a} polytropic index \revisefour{of} 0.5 \revisefour{and the compressive strength of dust aggregates (Equation (\ref{eq:comp}))}.

\item \revisethree{We discussed the general formation process of TNOs.
The lowest density TNO can be explained by primordial ice aggregates composed of 0.1-$\mathrm{\mu m}$-sized monomers, where primordial refers to those that have not undergone additional compaction.
The diversity of TNOs' bulk densities can be explained by the diversity of monomer radius and composition and by additional compaction due to melting or monomer disruption.}

\item We \revisethree{also} discussed the formation process of comets.
\revisethree{Even the lowest-density comet cannot be explained by primordial ice aggregates composed of 0.1--1.0 $\mathrm{\mu m}$-sized monomers.
If comets are primordial dust aggregates, the monomer radius should be} 1.0--10.0 $\mathrm{\mu m}$ for ice and 0.3--3.0 $\mathrm{\mu m}$ for silicate.
\revisethree{The comet formation process needs further growth modes: fragmentation and compaction.
In the case of fragmentation, which means comets are fragments or rubble piles of their parent dust aggregates, the diameters of parent bodies are $\sim60$ km or larger for 0.1-$\mathrm{\mu m}$-sized ice monomers, larger than 10 km for 0.1-$\mathrm{\mu m}$-sized silicate monomers, and $\sim6$ km or larger for 1.0-$\mathrm{\mu m}$-sized ice monomers.}

\item We also discussed the formation process \revisethree{of asteroids.}
\revisethree{MBAs can be explained by primordial silicate aggregates composed of 0.1--1.0 $\mathrm{\mu m}$-sized monomers.
The dispersion of MBAs' bulk densities can be explained by compaction due to melting or monomer disruption.
NEAs including Itokawa, Bennu, and Ryugu cannot be explained by primordial silicate aggregates composed of 0.1--1.0 $\mathrm{\mu m}$-sized monomers.
If NEAs are primordial silicate aggregates, the monomer radius should be $\gtrsim10.0\mathrm{\ \mu m}$, while Itokawa needs additional compaction.
The NEA formation process needs further growth modes: fragmentation and compaction.
In the case of fragmentation, the diameters of parent bodies are $\sim1.0\times10^2$ km or larger for 0.1-$\mathrm{\mu m}$-sized silicate monomers and $\sim10$ km for 1.0-$\mathrm{\mu m}$-sized silicate monomers.}
The \revisethree{previously suggested} diameter of the parent body of Ryugu, which is 100--160 km \citep{Walsh2013}, is consistent with our results of 0.1-$\mathrm{\mu m}$-sized silicate monomers.

\item We discussed the formation process of {\revisesix{observed compact dust aggregates}}.
{\revisesix{They}} can be explained by fragments of parent dust aggregates whose diameters are 30--180 km or larger for 0.1-$\mathrm{\mu m}$-sized ice monomers and 3--18 km or larger for 1.0-$\mathrm{\mu m}$-sized ice monomers.

\item We proposed a unified scenario to explain the formation process of planetesimals including \revisethree{TNOs,} comets, asteroids, and {\revisesix{observed compact dust aggregates}} in terms of direct coagulation as follows and shown in Figure \ref{fig:formation_scenario}.
First, 0.1-$\mathrm{\mu m}$-sized dust grains coagulate and grow into dust aggregates, and then the first-generation planetesimals composed of dust aggregates form.
The first-generation planetesimals include \revisethree{TNOs, MBAs, parent bodies of NEAs including Ryugu and Bennu, and parent bodies of comets.}
\revisefour{In this case, NEAs and comets are fragments or rubble piles of the first-generation planetesimals.}
Second, a part of the first-generation planetesimals fragment into {\revisesix{compact dust aggregates}}, they coagulate and grow into aggregates {\revisesix{of compact dust aggregates}}, and then the second-generation planetesimals composed of aggregates {\revisesix{of compact dust aggregates}} form.
The second-generation planetesimals \revisethree{include NEAs and comets.}

\end{enumerate}

\begin{acknowledgments}

We appreciate the discussion about monomer disruption with Sota Arakawa.
This work was supported by JSPS KAKENHI grant Nos. JP19J20351, JP22J00260, JP22KJ1292\revisetwo{, and JP22K03680}.
This work has made use of NASA’s Astrophysics Data System.
This work has made use of adstex (https://github.com/yymao/adstex).

\end{acknowledgments}

\appendix

\section{\revisefour{Accuracy of the Analytical Relation between Bulk Density and Diameter}}
\label{apsec:numerical}

\revisefour{In} this section, we numerically calculate the internal density and pressure profiles of dust aggregates by solving the hydrostatic equilibrium, and then we calculate their bulk density and compare it with the analytical \revisefour{relation} (Equation (\ref{eq:D-rho_bulk})).
\revisefour{The numerical calculation is necessary because the pressure, i.e., compressive strength, inside dust aggregates shows their polytropic index is not constant although the analytical relation is based on polytropes with a polytropic index of 0.5.}

First, we numerically solve the internal density and pressure profiles of dust aggregates.
The hydrostatic equilibrium is given as
\begin{equation}
\frac{d}{dr}\left(\frac{r^2}{\rho}\frac{dP}{d\rho}\frac{d\rho}{dr}\right) = -4\pi G\rho r^2,
\label{eq:LE}
\end{equation}
where $r$ is the radius from the center of the object and $P$ is the pressure. 
Here, the pressure is equivalent to the compressive strength of dust aggregates, and therefore we derive $dP/d\rho$ from Equation (\ref{eq:comp}) as
\begin{equation}
\frac{dP}{d\rho} = \frac{dP_\mathrm{comp}}{d\rho} = \frac{3E_\mathrm{roll}\rho_\mathrm{m}}{r_\mathrm{m}^3}\frac{1}{\rho^2}\left(\frac{\rho_\mathrm{m}}{\rho}-\frac{1}{\phi_\mathrm{max}}\right)^{-4}.
\label{eq:dPdrho}
\end{equation}

After numerically solving Equations (\ref{eq:LE}) and (\ref{eq:dPdrho}) by integrating from the center where $\rho=\rho_\mathrm{c}$ to the surface where $\rho=0$, we calculate the bulk density by dividing the total mass by its volume.

\begin{figure*}[ht!]
\plotone{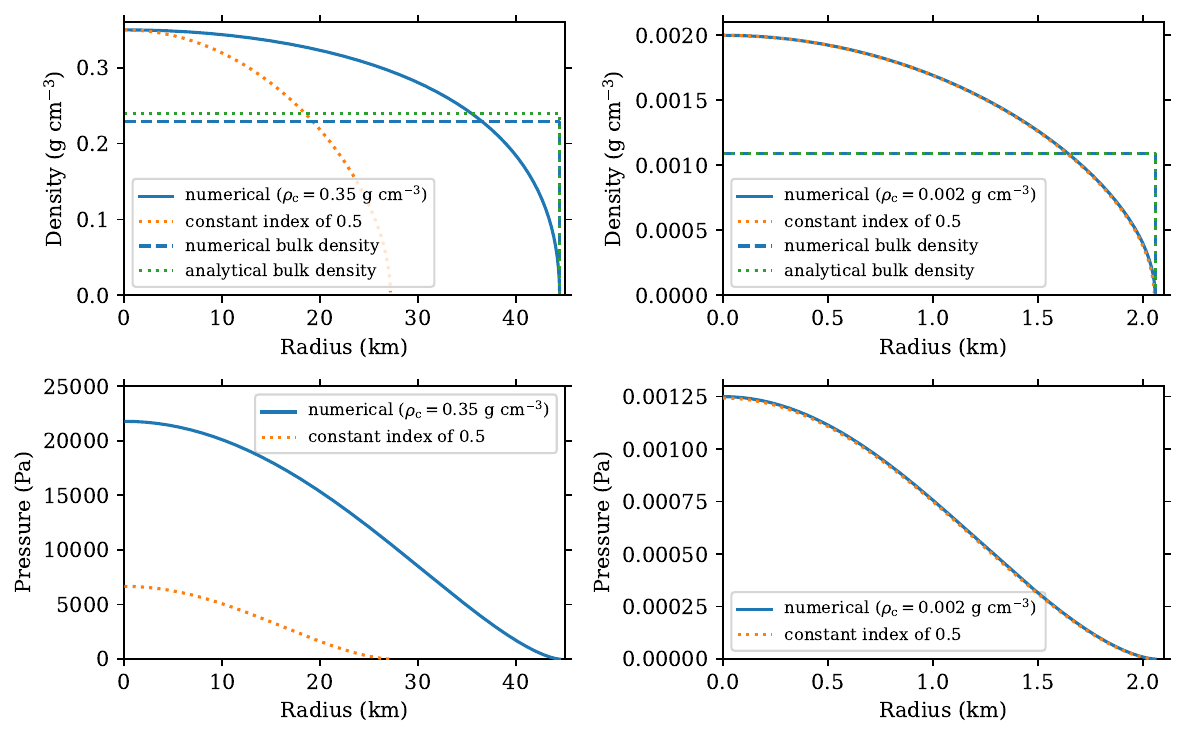}
\caption{\revisefour{Top: internal density profiles of ice dust aggregates (solid lines) when $r_\mathrm{m}=0.1\mathrm{\ \mu m}$ and $\rho_\mathrm{c}=0.35$ (left) and $0.002\mathrm{\ g\ cm^{-3}}$ (right).}
The other parameters but the material density are listed in Table \ref{tab:parameters}.
We assume that each ice monomer has a silicate core whose mass equals the mass of an ice mantle, so that $\rho_\mathrm{m}=1.45\mathrm{\ g\ cm^{-3}}$.
The \revisefour{orange} dotted lines show polytropes \revisefour{with a constant} polytropic index \revisefour{of} 0.5.
The dashed lines show the \revisefour{numerically obtained} bulk densities.
\revisefour{The green dotted lines show the bulk densities obtained analytically from Equation (\ref{eq:D-rho_bulk}).}
Bottom: internal pressure profiles.}
\label{fig:density-pressure-profile}
\end{figure*}

Figure \ref{fig:density-pressure-profile} shows the \revisefour{numerically} calculated internal density and pressure profiles of ice dust aggregates that have similar sizes to TNOs and comets.
The monomer radius is 0.1 $\mathrm{\mu m}$ and the other parameters but the material density are listed in Table \ref{tab:parameters}.
Here, we assume the central densities $\rho_\mathrm{c}=0.35$ and $0.002\mathrm{\ g\ cm^{-3}}$.
In these cases, the bulk densities $\simeq 0.23$ and $0.0011 \mathrm{\ g\ cm^{-3}}$, and the objects' radii $\simeq 44$ and 2.1 km, respectively (see the \revisefour{blue} dashed line in Figure \ref{fig:density-pressure-profile}).

For comparison, we also plot the internal density and pressure profiles of polytropes \revisefour{with a} \revisefour{constant} polytropic index \revisefour{of} 0.5 in Figure \ref{fig:density-pressure-profile}.
In the case of $\rho_\mathrm{c}=0.002\mathrm{\ g\ cm^{-3}}$, \revisefour{the} polytrope \revisefour{with a constant index agrees well with} the numerical results because the density is low enough, while \revisefour{in the case of $\rho_\mathrm{c}=0.35\mathrm{\ g\ cm^{-3}}$, the} polytrope \revisefour{with a constant index deviates from} the numerical results.
\revisefour{However, even in the latter case, the bulk density obtained from Equation (\ref{eq:D-rho_bulk}) agrees well with the numerically calculated bulk density.}

\begin{figure}[ht!]
\plotone{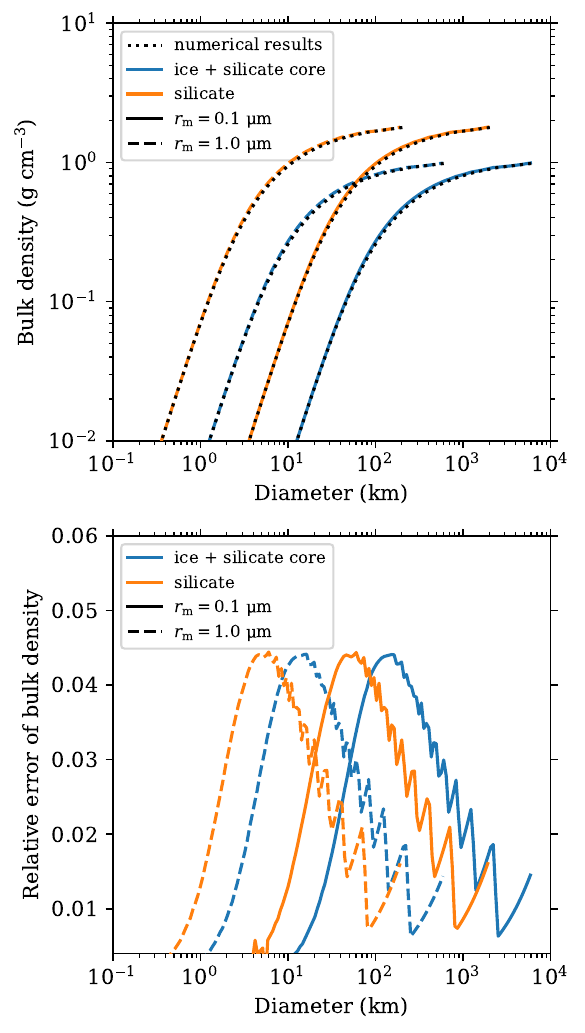}
\caption{\revisefour{Top: bulk density against diameter of dust aggregates.
The dotted lines show the numerically calculated bulk densities, while others show the analytical relation (Equation (\ref{eq:D-rho_bulk})).
The blue and orange lines show ice and silicate dust aggregates, whose material densities are 1.45 and 2.65 $\mathrm{g\ cm^{-3}}$, respectively.
The solid and dashed lines show monomer radii of 0.1 and 1.0 $\mathrm{\mu m}$, respectively.
Bottom:} relative error to the analytical \revisefour{relation} (Equation (\ref{eq:D-rho_bulk})) of numerically calculated bulk density.}
\label{fig:size-density-error}
\end{figure}

Figure \ref{fig:size-density-error} shows the relative error to the analytical \revisefour{relation} (Equation (\ref{eq:D-rho_bulk}))\revisefour{, which is based on polytropes with an index of 0.5 and the compressive strength of dust aggregates,} of numerically calculated bulk density.
The relative error is below 5\% for all cases, and therefore we conclude that the analytical relation corresponds to numerical calculated results.

\section{\revisefour{Central Properties of Dust Aggregates}}
\label{apsec:central}

In this section, we derive \revisefour{relations} for central density and central pressure against diameter to discuss the validity of our results in terms of central properties.

First, we derive a relation between central density $\rho_\mathrm{c}$ and diameter $D$ by using Equation (\ref{eq:polytrope}) as
\begin{eqnarray}
D &=& 2\left(\frac{3P_\mathrm{c}}{8\pi G\rho_\mathrm{c}^2}\right)^{1/2}\xi_1 \nonumber \\
&=& 2\left[\frac{3}{8\pi G}\frac{E_\mathrm{roll}}{r_\mathrm{m}^3}\left(\frac{\rho_\mathrm{m}}{\rho_\mathrm{c}}-\frac{1}{\phi_\mathrm{max}}\right)^{-3}\frac{1}{\rho_\mathrm{c}^2}\right]^{1/2}\xi_1 \nonumber\\
&\simeq& 3.89\left(\frac{E_\mathrm{roll}}{m_\mathrm{m}G\rho_\mathrm{m}}\right)^{1/2}\left(\frac{\rho_\mathrm{m}}{\rho_\mathrm{c}}-\frac{1}{\phi_\mathrm{max}}\right)^{-3/2}\frac{\rho_\mathrm{m}}{\rho_\mathrm{c}}.
\label{eq:D-rho_c}
\end{eqnarray}

Next, we derive a relation between central pressure $P_\mathrm{c}$ and diameter $D$.
By using Equation (\ref{eq:comp}), we obtain
\begin{eqnarray}
P_\mathrm{c} &=& \frac{E_\mathrm{roll}}{r_\mathrm{m}^3}\left(\frac{\rho_\mathrm{m}}{\rho_\mathrm{c}}-\frac{1}{\phi_\mathrm{max}}\right)^{-3},\\
\frac{\rho_\mathrm{m}}{\rho_\mathrm{c}} &=& \left(\frac{E_\mathrm{roll}}{P_\mathrm{c}r_\mathrm{m}^3}\right)^{1/3} + \frac{1}{\phi_\mathrm{max}}.
\label{eq:rho_m_rho_c}
\end{eqnarray}
We obtain the diameter $D$ by using Equations (\ref{eq:polytrope}) and (\ref{eq:rho_m_rho_c}) as
\begin{eqnarray}
D &=& 2\left(\frac{3P_\mathrm{c}}{8\pi G\rho_\mathrm{c}^2}\right)^{1/2}\xi_1 \nonumber \\
&=& 2\left\{\frac{3P_\mathrm{c}}{8\pi G}\frac{1}{\rho_\mathrm{m}^2}\left[\left(\frac{E_\mathrm{roll}}{P_\mathrm{c}r_\mathrm{m}^3}\right)^{1/3} + \frac{1}{\phi_\mathrm{max}}\right]^2\right\}^{1/2}\xi_1 \nonumber \\
&\simeq&1.90\left(\frac{P_\mathrm{c}}{G\rho_\mathrm{m}^2}\right)^{1/2}\left[\left(\frac{E_\mathrm{roll}}{P_\mathrm{c}r_\mathrm{m}^3}\right)^{1/3} + \frac{1}{\phi_\mathrm{max}}\right].
\label{eq:D-P_c}
\end{eqnarray}

\begin{figure}[b!]
\plotone{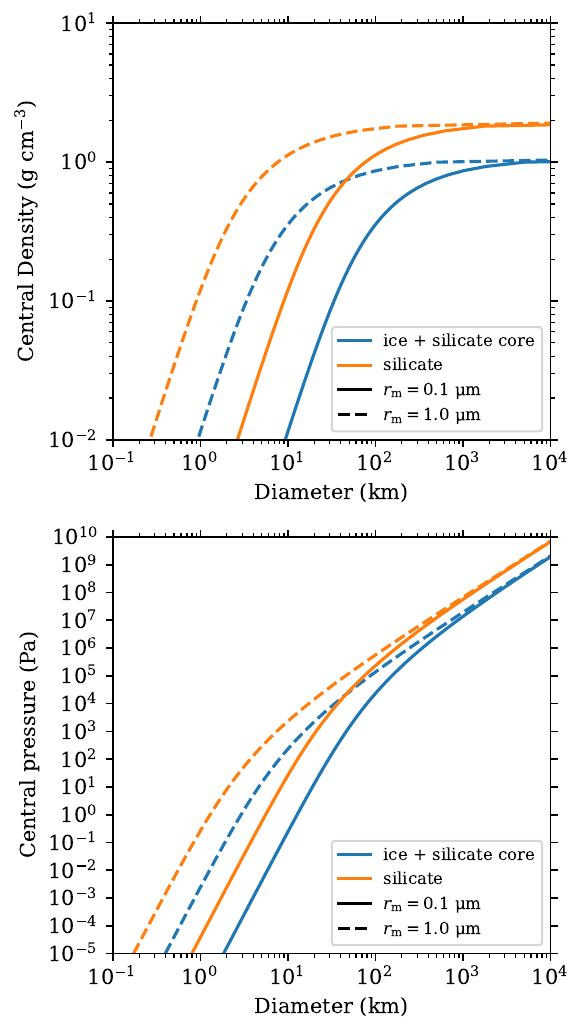} 
\caption{\revisefour{Top: central density against diameter of dust aggregates (Equation (\ref{eq:D-rho_c})).
The blue and orange lines show ice and silicate dust aggregates, whose material densities are 1.45 and 2.65 $\mathrm{g\ cm^{-3}}$, respectively.
The solid and dashed lines show monomer radii of 0.1 and 1.0 $\mathrm{\mu m}$, respectively.}
Bottom: central pressure against diameter of dust aggregates (Equation (\ref{eq:D-P_c})).
\label{fig:size-pressure}}
\end{figure}

Figure \ref{fig:size-pressure} shows the \revisefour{relations} for the central density (Equation (\ref{eq:D-rho_c})) and the central pressure (Equation (\ref{eq:D-P_c})) in four cases: 0.1-$\mathrm{\mu}$m-radius ice, 1.0-$\mathrm{\mu}$m-radius ice, 0.1-$\mathrm{\mu}$m-radius silicate, and 1.0-$\mathrm{\mu}$m-radius silicate monomers.

\section{\revisefour{Monomer Disruption}}
\label{apsec:monomer}

\revisethree{We discuss the validity of our results in terms of the pressure that monomers can be broken \citep[see Section 4.2 in][]{Tatsuuma2023}.
If the central pressure is too high, monomers can be broken, and then the bulk density and diameter can become different from our results.
The pressure at which materials can be broken $P_\mathrm{dis}$ has been investigated in the context of material science.
For example, ice can be broken at 5--25 MPa when the temperature is from $-10^\circ$C to $-20^\circ$C \citep{Haynes1978,Petrovic2003}.
Silica glasses, on the other hand, can be broken at $\sim5$ GPa at room temperature \citep{Proctor1967,Bruckner1970,Kurkjian2003}.
Here, we assume $P_\mathrm{dis}=10$ {\revisefive{MPa}} and 1 GPa for ice and silicate, respectively.
We note that these pressures are not for submicrometer--micrometer-sized spheres, so $P_\mathrm{dis}$ and further discussion can be different from our estimation.}

\begin{figure}[ht!]
\plotone{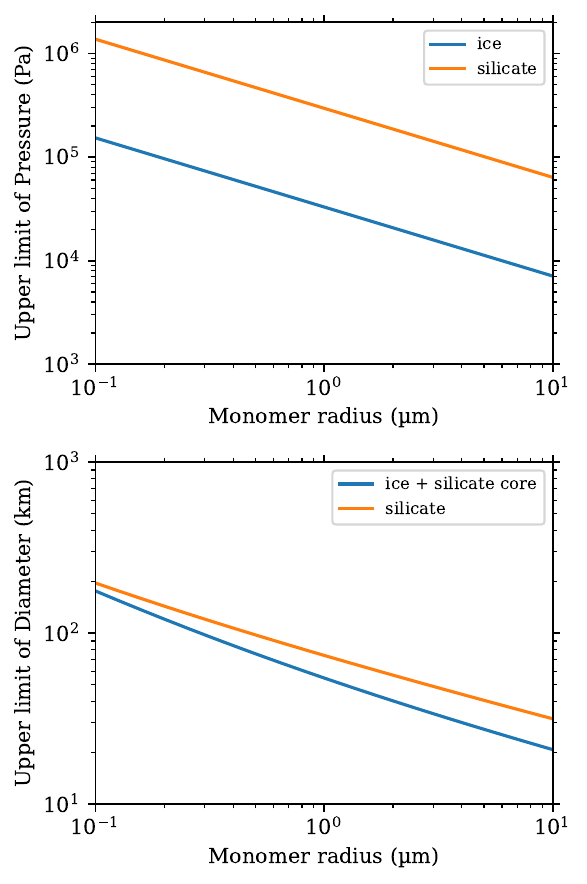} 
\caption{\revisefour{Top: upper limit of pressure (Equation (\ref{eq:P_upperlimit})) of dust aggregates below which monomer disruption can be neglected as a function of monomer radius.
The blue and orange lines are for ice and silicate, whose material densities are 1.45 and 2.65 $\mathrm{g\ cm^{-3}}$, respectively.
The parameters but the material density are listed in Table \ref{tab:parameters}.}
Bottom: upper limit of diameter of dust aggregates (Equation (\ref{eq:D_upperlimit})).
\label{fig:size_upper_limit}}
\end{figure}

We estimate the upper limit of diameter $D_\mathrm{ul}$ below which our results are valid according to the discussion in Section 4.2 in \citet{Tatsuuma2023}.
By considering the stress applied to the contact surface between monomers, we obtain the upper limit as
\begin{eqnarray}
P_\mathrm{ul} &\simeq& 0.014P_\mathrm{dis}\left(\frac{\gamma}{100\mathrm{\ mJ\ m^{-2}}}\right)^{2/3}\nonumber\\
&&\times\left(\frac{r_\mathrm{m}}{0.1\mathrm{\ \mu m}}\right)^{-2/3}\left(\frac{E^\ast}{3.7\mathrm{\ GPa}}\right)^{-2/3},\label{eq:P_upperlimit}\\
E^\ast &=& \frac{E}{2(1-\nu^2)},
\end{eqnarray}\\
\begin{equation}
D_\mathrm{ul} \simeq 1.90\left(\frac{P_\mathrm{ul}}{G\rho_\mathrm{m}^2}\right)^{1/2}\left[\left(\frac{E_\mathrm{roll}}{P_\mathrm{ul}r_\mathrm{m}^3}\right)^{1/3} + \frac{1}{\phi_\mathrm{max}}\right],
\label{eq:D_upperlimit}
\end{equation}
where $P_\mathrm{ul}$ is the upper limit of pressure, $E$ is the Young's modulus, $E^\ast$ is the reduced Young's modulus, and $\nu$ is the Poisson's ratio\revisefour{, by using Equation (\ref{eq:D-P_c}).}
We plot the upper limit of both pressure (Equation (\ref{eq:P_upperlimit})) and diameter (Equation (\ref{eq:D_upperlimit})) as a function of monomer radius in Figure \ref{fig:size_upper_limit}.
We find that the upper limit below which monomer disruption can be neglected is $D_\mathrm{ul} \sim 200$ km in the case of 0.1-$\mathrm{\mu m}$-radius monomers, which corresponds to the diameters of TNOs.

\bibliography{paper}{}
\bibliographystyle{aasjournal}

\end{document}